\documentstyle[pre,aps,epsfig,preprint]{revtex}
\begin{document}
\draft
\title{Stochastic resonance with different periodic forces in  overdamped \\two coupled
anharmonic oscillators}
\author{V.M.~Gandhimathi$^a$, K.~Murali$^b$, S.~Rajasekar$^{a,*}$}
\address{\vskip 5pt  $^{a}$School of Physics,
          Bharathidasan University, \\
         Tiruchirapalli--620 024,
         Tamilnadu, India
         \vskip 5pt  $^{b}$Department of Chemical Engineering,
          Anna University, \\
         Chennai--600 025,
         Tamilnadu, India}
\maketitle
 \vskip 425pt
 \hskip 50pt{$^*$Corresponding author. E-mail: rajasekar@physics.bdu.ac.in}
\newpage
\vskip 10 pt
\begin{abstract}
We study the stochastic resonance phenomenon in the overdamped two
coupled anharmonic oscillators with Gaussian noise and driven by
different external periodic forces. We consider (i) sine, (ii)
square, (iii) symmetric saw-tooth, (iv) asymmetric saw-tooth, (v)
modulus of sine and (vi) rectified sinusoidal forces. The external
periodic forces and Gaussian noise term are added to one of the
two state variables of the system. The effect of each force is
studied separately. In the absence of noise term, when the
amplitude $f$ of the applied periodic force is varied cross-well
motion is realized above a critical value ($f_{\mathrm{c}}$) of
$f$. This is found for all the forces except the modulus of sine
and rectified sinusoidal forces. For fixed values of angular
frequency $\omega$ of the periodic forces, $f_{\mathrm{c}}$ is
minimum for square wave and maximum for asymmetric saw-tooth wave.
$f_{\mathrm{c}}$ is found to scale as $ A {\mathrm{e}^{0.75
\omega}} + B$ where $A$ and $B$ are constants. Stochastic
resonance is observed in the presence of noise and periodic
forces. The effect of different forces is compared. The stochastic
resonance behaviour is quantized using power spectrum,
signal-to-noise ratio, mean residence time and distribution of
normalized residence times. The logarithmic plot of mean residence
time $\tau_{\mathrm{MR}}$ against $ 1/(D - D_{\mathrm{c}})$ where
$D$ is the intensity of the noise and $D_{\mathrm{c}}$ is the
value of $D$ at which cross-well motion is initiated shows a sharp
knee-like structure for all the forces. Signal-to-noise ratio is
found to be maximum at the noise intensity $D=D_{\mathrm{max}}$ at
which mean residence time is half of the period of the driving
force for the forces such as sine, square, symmetric saw-tooth and
asymmetric saw-tooth waves. With modulus of sine wave and
rectified sine wave, the $SNR$ peaks at a value of $D$ for which
sum of $\tau_{MR}$ in two wells of the potential of the system is
half of the period of the driving force. For the chosen values of
$f$ and $\omega$, signal-to-noise ratio is found to be maximum for
square wave while it is minimum for modulus of sine and rectified
sinusoidal waves. The values of $D_{\mathrm{c}}$ at which
cross-well behaviour is initiated and $D_{\mathrm{max}}$ are found
to depend on the shape of the periodic forces.
\end{abstract}
 \vskip 10pt
\pacs{{\emph{PACS}}: 02.50.-r; 05.40.-a; 05.45.-a  \\
\noindent{{\emph{Keywords}: Over damped two coupled anharmonic
oscillators; stochastic resonance; periodic forces;
 power spectrum, mean residence time;  signal-to-noise ratio.
  }} }
 \newpage
\noindent {\bf{1. Introduction}}
 \vskip 10pt
 The phenomenon of stochastic resonance has simulated series of
 theoretical, numerical and experimental works on the dynamics of
 multi-stable systems subjected to both periodic and noisy
 driving. In many nonlinear systems, the occurrence of stochastic
 resonance has been studied with the external periodic force being
 of the form $f \sin \omega t$ [1-16]. There are few studies with
 aperiodic forces. For example, enhancement of response of certain
 nonlinear systems by the added noise with aperiodic forces is
 observed [17-20]. Jamming (undesirable noise) occurring in an
 array of FitzHugh-Nagumo oscillators was found to be suppressed
 via stochastic resonance by the input aperiodic signals such as
 weak amplitude modulated, frequency modulated and chaotic
 signals [21]. Stochastic resonance can happen in the absence of
 periodic excitation [5, 22] and noise [23, 24] also. The former was
 called spontaneous stochastic resonance and latter as
 deterministic stochastic resonance.
 \vskip 10 pt
 In recent years, the dynamics of nonlinear systems with different
 periodic forces has been studied [25-30]. Particularly, the onset
 of homoclinic chaos in a damped pendulum driven by a periodic
 string of pulses modulated by Jacobian elliptic function and
 periodic $\delta$-function is investigated [25]. The existence of a
 new multiple-period-doubling bifurcation route to chaos and
 modified bifurcation structures in the Duffing oscillator
 perturbed by periodic pulses are reported [26]. Anti-control of
 chaos is accomplished by adding certain periodic forces [27]. A
 nonlinear feedback controller that generates chaotic behaviour in
 an oscillator driven by a distorted force has been proposed [28].
 Suppression of chaos by $\delta$-pulse in Duffing-van der Pol
 oscillator [29] and by a variable shape pulse function in a
 coupled pendulum-harmonic oscillator system [30] is found.
 \vskip 10 pt
 Certain dynamics are studied with forces other than $f \sin \omega
 t$. It is important to study a particular dynamics with different
 types of forces and make a comparative study of effects induced
 by them. This is because different forms of periodic forces can
 be generated and easily applied to real mechanical systems and
 electronic circuits. Further, the study of influence of various
 forces in nonlinear systems will be helpful to choose a suitable
 force in creating and controlling nonlinear behaviours.
 \vskip 10 pt
 Motivated by the above, in the present paper, we wish to study
 stochastic resonance phenomenon with different periodic forces in
 the overdamped two coupled anharmonic oscillators given by [31]

\begin{mathletters}
\begin{eqnarray}
\dot {x} & = & a_1 x - b_1 x^3 + \delta xy^2 + F(t)
               +  \eta(t), \\
\dot {y} & = & a_2 y - b_2 y^3 + \delta x^2y.
\end{eqnarray}
\end{mathletters}
In eqs(1) $F(t)$  and  $\eta(t)$ represent the external forcing
term and the noise term respectively. In the absence of external
periodic force, noise and damping terms the potential of the two
coupled anharmonic oscillators is
\begin{equation}
V(x,y)  =  -  \frac{a_1}{2}x^2 + \frac{b_1}{4}x^4 - \frac{a_2}{2}
    y^2 + \frac{b_2}{4}y^4 - \frac{\delta}{2}x^2y^2.
\end{equation}
In a very recent work, Baxter et al [32] considered the eqs(1) as
a model for competition between two species. They analysed the
probability of survival of the species using the path integral
method in eqs(1) without $F(t)$ and adding noise to both the state
varaibles $x$ and $y$. Different types of intermittent-lag
synchronization are shown to occur in a parametrically driven two
coupled anharmonic oscillators [33].

 \vskip 10pt
The organization of the paper is as follows. In section 2, to be
self-content we give the mathematical form of the different
external forces used in our study. The effect of periodic forces
without noise term is discussed in section 3.  For fixed values of
the parameters, the system is found to exhibit cross-well periodic
orbit above a certain critical value ($f_{\mathrm{c}}$) of the
amplitude of the forces. The value of $f_{\mathrm{c}}$ depends on
the shape of the forces. $f_{\mathrm{c}}$ is found to increase
with increase in the frequency $ \omega$ of the forces and scales
as $A {\mathrm{e}^{0.75 \omega}} + B$. In section 4 we consider
the system in the presence of external periodic force and noise.
For each force except the modulus of sine and rectified sinusoidal
wave stochastic resonance is observed at $D_{\mathrm{max}}$ at
which mean residence time is half of the period of the driving
force. Stochastic resonance is characterized using power spectrum,
signal-to-noise ratio, mean residence time and distribution of
normalized residence time. We compare the effect of various forces
on stochastic resonance phenomenon in terms of its characteristic
quantities. Section 5 contains summary and conclusion of our
present study.
 \vskip 10pt
\noindent{\bf{2. Types of periodic forces}}
 \vskip 10pt
The periodic forces of our interest are \vskip 1 pt
 \hskip 3 pt 1. sine wave \vskip 1 pt
 \hskip 3 pt 2. square wave \vskip 1 pt
 \hskip 3 pt 3. symmetric saw-tooth \vskip 1 pt
 \hskip 3 pt 4. asymmetric saw-tooth \vskip 1 pt
 \hskip 3 pt 5. modulus of sine wave \vskip 1 pt
 \hskip 3 pt 6. rectified sine wave \vskip 1 pt

\noindent Figure (1) shows the shape of the above forces.
Throughout our study, we fix the period $T_{0}$ of all the forces
as $ 2 \pi / \omega$. \vskip 2 pt
 \hskip 20 pt The mathematical representations of the periodic forces are the
 following:
  \vskip 3pt
\noindent {\bf(1) Sine wave} \\
 The sine wave is given by the form
 \begin{equation}
 F_{\mbox{si}}(t) = f \sin \omega t,
 \end{equation}
 where $f$ is the amplitude of the force and $\omega$ is the
 angular frequency of the force.
 \vskip 3pt \noindent {\bf (2) Square wave} \\
 The mathematical form of square wave is
 \begin{equation}
 F_{\mbox{sq}}(t) =
 \left\{\begin{array}{ll}
                 f, \quad (2n-2) \pi/\omega < t < (2n-1)
\pi/\omega \\
 -f, \quad (2n-1) \pi/\omega < t < 2n
 \pi/\omega, \quad n = 1,2, \ldots
\end{array}\right.
 \end{equation}
For numerical implementation, we rewrite the force (4) as
 \begin{equation}
 F_{\mbox{sq}}(t) = \left\{\begin{array}{ll}
   f, \quad 0 < t < {\pi}/{\omega} \\
 -f,\quad {\pi}/{\omega}< t < {2 \pi}/{\omega},
 \end{array}\right.
 \end{equation}
 where $t$ is taken as mod($ 2 \pi / \omega$). That is,
$F_{\mbox{sq}}(t)$ = $F_{\mbox{sq}}(t +{2 \pi}/{\omega})$. Its
Fourier series is
\begin{equation}
F_{\mbox{sq}}(t) = \frac {4f} {\pi} \sum^{\infty}_{n = 1}
\frac{\sin (2n-1) \omega t}{(2n-1)}.
\end{equation}
\vskip 3 pt \noindent {\bf(3) Symmetric saw-tooth wave} \\
  Symmetric saw-tooth wave is mathematically represented as
\begin{equation}
 F_{\mbox{sst}}(t) = \left\{\begin{array}{ll}  \frac {4ft} {T_{0}}, \quad
 (2n-2) {\pi}/{\omega} < t < \left (\frac
 {4n-3}{2}\right) {\pi}/{\omega}\\
 - \frac{4ft} {T_{0}} + 2f, \quad \left
(\frac {4n-3}{2}\right) {\pi}/{\omega} < t < \left (\frac
 {4n-1}{2}\right) {\pi}/{\omega} \\
 \frac{4ft} {T_{0}} - 4f, \quad \left (\frac {4n-1}{2}\right)
{\pi}/{\omega} < t < 2n {\pi}/{\omega}, \quad n =1,2, \ldots
 \end{array}\right.
 \end{equation}
 where $T_{0}$ is the period of the force. The numerical
 implementation of the force is given by
 \begin{equation}
  F_{\mbox{sst}}(t) = \left\{\begin{array}{ll}
  \frac{4ft} {T_{0}}, \quad 0 < t < {\pi}/{2 \omega}  \\
 - \frac{4ft} {T_{0}} + 2f, \quad {\pi}/{2 \omega} < t <
  {3 \pi}/{2 \omega} \\
 \frac{4ft} {T_{0}} - 4f, \quad {3 \pi}/{2 \omega} < t <
  {2 \pi}/{ \omega},
  \end{array}\right.
  \end{equation}
 where $t$ is taken as mod($ 2 \pi / \omega$). Its Fourier series is
\begin{equation}
F_{\mbox{sst}}(t) = \frac {8f} {\pi^{2}} \sum^{\infty}_{n =
1} \frac{(-1)^{(n+1)} \sin (2n-1) \omega t}{(2n-1)^{2}}.
\end{equation}
\vskip 3 pt
 \noindent {\bf{(4) Asymmetric saw-tooth wave}} \\
  The mathematical representation of the wave form shown in
  fig.1d is
 \begin{equation} F_{\mbox{ast}}(t) = \left\{\begin{array}{ll}
  \frac{2ft} {T_{0}}, \quad  (2n-2){\pi}/{\omega} < t <
 (2n-1) {\pi}/{\omega} \\
 \frac{2ft} {T_{0}} - 2f, \quad (2n-1) {\pi}/{\omega} < t < 2n
{\pi}/{\omega},  \quad n = 1,2, \dots
 \end{array}\right.
 \end{equation}
 The force (10) is rewritten as
 \begin{equation}
  F_{\mbox{ast}}(t) = \left\{\begin{array}{ll} \frac{2ft} {T_{0}},
  \quad 0 < t < {\pi}/{\omega} \\
  \frac{2ft} {T_{0}} - 2f, \quad {\pi}/{\omega} < t <
  {2 \pi}/{ \omega},
  \end{array}\right.
  \end{equation}
 where $t$ is taken as mod($ 2 \pi / \omega$). The Fourier series of $F_{\mbox{ast}}(t)$ is given by
\begin{equation}F_{\mbox{ast}}(t) = \frac {2f} {\pi} \sum^{\infty}_{n = 1}
\frac{(-1)^{(n+1)}  \sin n \omega t}{n}.
\end{equation}
\vskip 3 pt \noindent {\bf{(5) Modulus of sine wave}} \\ The
modulus of sine wave is given as
 \begin{equation}
F_{\mbox{msi}}(t) = f \vert \sin  ({\omega t}/{2})  \vert.
\end{equation}
 Its Fourier series is
\begin{equation}
F_{\mbox{msi}}(t) =  \frac {2f} {\pi} - \frac {4f} {\pi} \sum^{\infty}_{n = 1}
\frac{ \cos n \omega t}{(4n^{2}-1)}.
\end{equation}
\vskip 3 pt
 \noindent {\bf{(6) Rectified sine wave}}\\
The mathematical representation of rectified sine wave (fig.1f) is
\begin{equation}
 F_{\mbox{rsi}}(t) = \left\{\begin{array}{ll}f, \quad  (2n-2) {\pi}/{\omega}
 < t < (2n-1) {\pi}/{\omega} \\
 0, \quad (2n-1) {\pi}/{\omega} < t < 2n {\pi}/{\omega}, \quad n = 1,2,\dots
 \end{array}\right.
 \end{equation}
 The numerical implementation of the force (15) is
 \begin{equation}
  F_{
  \mbox{rsi}}(t) = \left\{\begin{array}{ll} f, \quad 0 < t <
  {\pi}/{\omega} \\
 0, \quad {\pi}/{\omega} < t < {2 \pi}/{\omega}.
  \end{array}\right.
  \end{equation}
 The Fourier series of rectified sine wave is
 \begin{equation} F_{\mbox{rsi}}(t) =  \frac {1} {\pi} - \frac {2} {\pi} \sum^{\infty}_{n = 1}
  \frac  {\cos 2n \omega t} {(4n^{2}-1)} + \frac {\sin \omega t} {2}.
\end{equation}
for $0 < t < \pi/ \omega$ and $0$ for $\pi/ \omega < t <  2 \pi/ \omega$.
 \vskip 10 pt
\noindent{\bf{3. Cross-well motion due to different periodic
forces without noise term}}
 \vskip 10 pt
 In this section we considered the system (1) in the absence of
 noise term $\eta(t)$ but with different external forces. The
 parameters of the system are fixed at $ a_1 = 1.0$, $a_2 = 1.1$, $b_1 = 1.0$, $b_2 =
 1.0$. For this choice, the potential is a four-well potential.
 The potential wells are designated as $V_{++}$ for $x>0$, $y>0$; $V_{+-}$ for $x>0$,
 $y<0$; $V_{-+}$ for $x<0$, $y>0$; $V_{--}$ for $x<0$, $y<0$. The
 period of the all the forces is set to $ 2 \pi / \omega$. For
 each force the response of the system is studied by varying the
 amplitude of it.
 \vskip 10 pt
 First, we consider the effect of the sinusoidal force $ f \sin
 \omega t$. For a fixed value of $\omega$ and for small values of
 $f$, four period-$T_{0}$ ($ 2 \pi / \omega$) orbits one in each
 potential well exist. The size of the orbit increases with
 increase in the value of $f$. At a critical value of $f$,
 $f_{\mathrm{c}}$, the trajectory makes a cross-well motion forming
 a coupled orbit. However, the period of the cross-well orbit remains as
 $T_{0}$. Since the external forcing is added only to
$x$-component of the system, the coupled orbit traverse the wells
$V_{++}$ and $V_{-+}$ or the wells $V_{+-}$ and $V_{--}$ depending
upon the initial conditions. For $f$ in the range [0, 50], chaotic
behaviour is not observed. This is confirmed from the bifurcation
diagram and Lyapunov exponent calculations. Similar behaviour is
observed for various values of $ \omega$ in the interval [0, 1]
with step size $ \omega = 0.05$ and also for other forces except
for modulus of sine wave and rectified sine wave. In the case of
modulus of sine wave added to $x$-component of the system the
wells $V_{-+}$ and $V_{--}$ alone are oscillating. The wells
$V_{++}$ and $V_{+-}$ are stationary. For initial conditions in
the well $V_{+-}$ or $V_{--}$, after the transient evolution the
system remains in the well $V_{+-}$. In the case of rectified sine
wave during the first half of the period of the drive cycle only
the wells $V_{-+}$ and $V_{--}$ oscillate. Therefore, cross-well
behaviour is not observed for $f$ in the range [0, 50] in both the
cases.
 \vskip 10 pt
For the force $f \sin \omega t$ with $ \omega = 0.05$, the
critical value of $f$ (at which cross-well motion is initiated) is
numerically found to be $0.409$. For forces such as square,
symmetric saw-tooth and asymmetric saw-tooth waves
$f_{\mathrm{c}}$ is $0.393$, $0.4408$ and $0.475$ respectively.
Figure 2 shows the variation of $f_{\mathrm{c}}$ with $\omega$. In
this figure the dots represent numerical result and the continuous
lines are the best curve fit. We find that $f_{\mathrm{c}}$
increases with increase in $\omega$. For a fixed $\omega$ the
$f_{\mathrm{c}}$ for square wave is minimum while it is maximum
for asymmetric saw-tooth wave. That is, cross-well motion is
initiated at a lower value of $f$ for square wave when compared
with other forces. The cross-well motion occurs relatively at a
higher value of $f$ in the case of sine, symmetric saw-tooth and
asymmetric saw-tooth waves. For all the forces $f_{\mathrm{c}}$ is
found to vary with $\omega$ as $A {\mathrm{e}^{0.75 \omega}} + B$.
The values of $(A,B)$ obtained for square, sine, symmetric
saw-tooth and asymmetric saw-tooth forces are $(0.471, -0.130)$,
$(0.754, -0.409)$, $(0.987, -0.602)$ and $(1.284, -0.834)$
respectively.
 \vskip10pt
\noindent{\bf{4. Stochastic resonance with different periodic
forces}}
 \vskip 10 pt
In the previous section, we studied the effect of different
periodic forces separately in the absence of noise term $
\eta(t)$. Now, we include the noise term and study the stochastic
resonance phenomenon by varying the noise intensity $D$. Here
again we perform the analysis for each force separately. The noise
term is chosen as Gaussian random numbers. The parameters are
fixed at $ a_1 = 1.0$, $a_2 = 1.1$, $b_1 = 1.0$, $b_2 = 1.0$,
$\delta = 0.01$ and $\omega = 0.05$. In the absence of noise, the
critical value of $f$ at which cross-well orbit occurs for square,
sine, symmetric saw-tooth and asymmetric saw-tooth waves are $
0.393$, $0.409$, $0.4408$ and $0.475$ respectively. The value of
$f$ is fixed below $f_{\mathrm{c}}$. We fix $ f = 0.38$ for all
the forces so that in the absence of noise the motion is confined
to single well alone, that is, there is no cross-well motion. We
integrated eqs(1) using a fourth-order Runge-Kutta method with
step size $\Delta t = (2\pi/\omega)/N$, $N = 2001$. Noise is added
to the state variable $x$ as $ x_{i+1} \rightarrow x_{i+1} +
\sqrt{\Delta t}  D  \zeta(t) $ after each integration step $\Delta
t $. Here $ \zeta(t) $ represents Gaussian random numbers with
zero mean and unit variance and $D$ denotes noise intensity.
\vskip10pt \noindent{\bf{4.1. Cross-well motion and mean residence
time}}
 \vskip 10 pt
First, we illustrate the effect of noise in the presence of square
wave. Cross-well motion is not realized for $D < D_{\mathrm{c}}
\approx 0.0002$. At $D = D_{\mathrm{c}}$, orbit switching from
one-well to another well is initiated. For example, cross-well
behaviour is observed between the two wells $V_{++}$ and $V_{-+}$
at $D = D_{\mathrm{c}}$. Similar behaviour is observed when the
square wave is replaced by other wave forms with same $\omega$ and
$f$ values. However, the value of $D_{\mathrm{c}}$ is found to be
different for other forces. $ D_{\mathrm{c}}$ values for square,
sine, symmetric saw-tooth, asymmetric saw-tooth, modulus of sine
and rectified sine waves are $0.0002$, $0.0005$, $0.002$,
$0.0025$, $0.05$ and $0.03$ respectively. $D_{\mathrm{c}}$ is
relatively low for the square wave.
 \vskip 10 pt
For each periodic force, the switching between the wells become
very rare for $D$ values just above $D_{\mathrm{c}}$. However, as
$D$ increases switching also increases. To characterize this
behaviour we numerically calculated the mean residence time.
Residence time $\tau_{R}$ is defined as the duration of the
trajectory of the system residing in a well (for example,
$V_{++}$) before jumping to another well (say, $V_{-+}$) and
vice-versa. Mean residence time $\tau_{MR}$ in the wells $V_{-+}$
and $V_{++}$ are computed for a set of $10^5$ residence times.
Rapid variation of $\tau_{MR}$ is found when the noise intensity
$D$ is varied from $D_{\mathrm{c}}$. Figure 3a shows
$\ln(\tau_{\mathrm{MR}})$ versus $ 1/(D - D_{\mathrm{c}})$ when
the applied force is square wave. Later, we will show that signal-to-noise
ratio ($SNR$) is maximum at this value of $D$. Different form of
variation of $\tau_{MR}$ is found for $D < D_{\mathrm{max}}$ and
$D > D_{\mathrm{max}}$. The curve in the fig.3a has a sharp knee
at $D = D_{\mathrm{max}}$. At this value of $D$ the derivative of
$\tau_{MR}$ is discontinuous. Sharp-knee shape like variation of
maximal Lyapunov exponent versus control parameter is observed for
band-merging crisis [34]. Mean residence time $\tau_{MR}$ for the
other periodic forces are also calculated. The logarithmic plot of
$\tau_{MR}$ versus $ 1/(D - D_{\mathrm{c}})$ is shown in figs.3b-d
for sine, symmetric saw-tooth and asymmetric saw-tooth forces. All
the plots resemble the knee-like structure. The values of
$D_{\mathrm{max}}$ are found to be $0.14$, $0.1$, $0.12$ and
$0.17$ for square, sine, symmetric saw-tooth and asymmetric
saw-tooth forces.  The maximum $SNR$ values are obtained at these
values of $D$ for the above forces.
 \vskip 10 pt
For the forces depicted in figs.1a-d the sign of the forces is
positive during one half of the cycle and negative during the
second half of the cycle. For the forces in fig.1e and 1f the sign
of the force is never negative. This difference in the forces and
the symmetry between the wells have influence on the mean
residence time. For the forces shown in figs.1a-d for $D$ values
just above $D_{\mathrm{max}}$ the $\tau_{MR}$ in the wells
$V_{-+}$ and $V_{++}$ are same and greater than the period $T_{0}$
= $2 \pi / \omega \approx 125.67$ of the drive cycle. As $D$
increases $\tau_{MR}$ decreases. In the case of modulus of sine
and rectified sine waves, the mean residence time $\tau_{MR}$ in
the well $V_{-+}$ for $D$ just above $D_{\mathrm{c}}$ is less than
$ T_{0}/2 = \pi / \omega \approx 62.83$ while it is greater than
$T_{0}$ for the other well $V_{++}$. That is, $\tau_{MR}$ is not
same for the two wells $V_{-+}$ and $V_{++}$. $\tau_{MR}$ in both
the wells decreases with further increase in $D$. Figures 4 and 5
show the variation of $\tau_{MR}$ with $D$ for the modulus of sine
and rectified sine waves respectively.
 \vskip10pt \noindent{\bf{4.2. Stochastic resonance}}
 \vskip 10 pt Time series plot in the presence
of external square wave for few values of noise intensity $D$ is
shown in fig.6. For small values of $D$ the system exhibit the
same periodic behaviour of noise free case but slightly perturbed
by the noise. For $D < D_{\mathrm{c}} = 0.0002$, the trajectory of
the system resides in a single well as in fig.6a. At the critical
value $D_{\mathrm{c}}$ the trajectory jumps randomly from one well
to another. In fig.6b for $D = 0.001$ just above $D_{\mathrm{c}}$
the state variable $x$ switches irregularly and rarely between the
wells $V_{++}$ and $V_{-+}$. In the presence of forcing, the
system initially in the well, say, $V_{++}$ is forced by the noise
to leave the well. Then the system enters the well $V_{-+}$ and
wanders irregularly there for some time and jumps back to the well
$V_{++}$ and so on. For $D = 0.001$ $\tau_{MR}$ is computed as $
\approx 790$. The switching is not periodic. The residence times
are randomly distributed. $\tau_{MR}$ decreases with increase in
$D$. For $D = 0.14$, $\tau_{MR}$ is found to be $\approx \pi/
\omega$ $\approx 62.83$. In this case nearly periodic switching
between the wells $V_{++}$ and $V_{-+}$ occur. There is a
co-operation between the periodic driving force and the noise. The
$x$-component of the trajectory switches between the positive and
negative values with the period approximately half of the period
of the applied external periodic force. This is clearly seen in
fig.6c. \vskip 10 pt For large values of $D$, the motion is
dominated by the noise term. Figure 6d shows the trajectory for $D
= 2.5$ where the trajectory jumps erratically between the wells.
That is, loss of coherence is produced by the large noise
intensity. For $D$ in the range [$D_{\mathrm{c}}$, 4], jumping
motion between the wells $V_{++}$ and $V_{-+}$ alone observed. The
system (1) is studied with different periodic forces as a function
of noise intensity. We observed similar response of the system for
all the forces except modulus of sine and rectified sine waves.
\vskip 10 pt The trajectory of the system exhibited periodic
transitions between the wells at $D_{\mathrm{max}} = 0.1$, $0.12$,
$0.17$ for sine, symmetric saw-tooth and asymmetric saw-tooth
waves. Synchronization of the input wave form with the output at
these values of $D$  is shown in fig.7 for various forces. In this
figure for all the forces at $D = D_{\mathrm{max}}$ we can notice
a common feature. Nearly at the end of one half of a drive cycle
the trajectory in one well is likely to jump to the another well
and after the next half cycle it is likely to return back. This is
the signature of stochastic resonance.
\vskip10pt
\noindent{\bf{4.3. Characterization of stochastic resonance by
signal-to-noise ratio}}
 \vskip 10 pt
We characterize the stochastic resonance phenomenon in terms of
$SNR$ and the peak value at the half driving frequency in the
distribution of normalized residence times. $SNR$ is calculated
from the power spectrum of the state variable $x$. To compute
power spectrum, we have used fast Fourier transform technique [35,
36]. The power spectrum is obtained using a set of $2^{10}$ data
collected at a time interval of $(2 \pi/ \omega)/10$. To get more
accurate power spectrum we computed it for $25$ different
realization of Gaussian random numbers and then obtained the
average spectrum.
 \vskip 10 pt
In the absence of noise term, the power spectrum of $x$ component
of the system with different periodic forces is first analyzed.
With square wave as the external forcing, the power spectrum has
peaks at few odd and even integral multiples of the forcing
frequency. When the noise is included, the peaks at odd integral
multiples are alone dominant. With increase in noise intensity
$D$, the height of the peaks increase up to $D = D_{\mathrm{max}}$
(at which $ \tau_{\mathrm{MR}} \approx \pi / \omega$) and then
decrese. The power spectrum for four values of $D$ in the presence
of square wave is shown in fig.8. Dominant peaks are seen at $
\Omega = 0.05$ and $0.15$.
 \vskip 10 pt
 Signal-to-noise ratio is calculated from power
 spectrum using the formula
\begin{equation}
  SNR = 10 \log_{10}(S/N) \,\, \mathrm{dB}.
  \end{equation}
 In eq(18) $S$ and $N$ are the amplitudes of the signal peak and the noise
 background respectively. $S$ is read directly from the power spectrum
 at the frequency $\omega$ of the driving periodic force. To calculate
 the background of the power spectrum about $\omega$ we considered the power
 spectrum in the interval $\Omega  = \omega - \Delta \omega$ and
$\omega + \Delta \omega$ after subtracting the point namely
$\Omega = \omega$ representing the stochastic resonance spike. The
average value of the power spectrum in the above interval is taken
as background noise level at $\omega$. We have chosen $\Delta
\omega = 0.01$. Figure 9a shows the plot of $SNR$ as a function of
$D$ for square wave. The calculated $SNR$ increases rapidly with
the noise intensity $D$, peaks at the critical value
$D_{\mathrm{max}} = 0.14$ and then decreases slowly for $ D
> D_{\mathrm{max}}$. With modulus of sine and rectified sine waves for small values of
noise intensity $D$, rare switching between the wells $V_{++}$ and
$V_{-+}$ is observed. However, periodic switching with the half of
the period of the drive cycle is not observed when $D$ is varied.
Figure 9b shows $SNR$ versus $D$ for various applied forces. With
the forces such as sine, square, symmetric saw-tooth and
asymmetric saw-tooth waves, $SNR$ has a peak at $D =
D_{\mathrm{max}}$ at which $\tau_{\mathrm{MR}} \approx \pi/\omega$
and there is almost a periodic switching between the wells
$V_{-+}$ and $V_{++}$ as shown in fig.7. For the forces given in
figs.1e-f the $\tau_{\mathrm{MR}}$ in $V_{-+}$ is always less than
$\pi/\omega$. However, stochastic resonance is observed at the
value of $D$ at which the sum of the mean residence times of both
the wells $V_{-+}$ and $V_{++}$ is $ \approx \pi / \omega$. This
happens at $D =0.35$ and $D=0.3$ for modulus of sine and rectified
sine waves respectively. For these forces $SNR$ initially
decreases with increase in $D$ from $D_{\mathrm{c}}$ for a while,
increases with increase in $D$, reaches a maximum value and then
decreases with further increase in $D$. $SNR$ is found to be
highest for square wave. Modulus of sine wave and rectified sine
wave yields the minimum $SNR$ value.
 \vskip10pt
 \noindent{\bf{4.4. Characterization of stochastic resonance by probability
distribution of normalized residence times}}
 \vskip 10 pt The probability distribution of normalized
residence times denoted as $P$ can also be used to quantify
stochastic resonance. Normalized residence time is calculated as
follows. For a fixed noise intensity $D$ the residence time
$\tau_{R}$ in a well is computed for a set of $10^{5}$
transitions. Then, normalized residence times are obtained by
dividing $\tau_{R}$ by the period $T_{0}$ of the applied force.
Then we calculated $P(\tau_{R}/T_{0})$. $P$ is found to be same
for the wells $V_{-+}$ and $V_{-+}$. In fig.10 we have plotted $P$
versus $\tau_{R}/T_{0}$ for four values of noise intensity with
the applied periodic force being square wave. We observe a series
of Gaussian like peaks, centered at odd integral multiples of
$T_{0}/2$. The heights of these peaks decrease exponentially with
their order $n$. The strength $P_{1}$ of the peak at $\tau_{R} =
T_{0}/2$ is a measure of the synchronization between the periodic
forcing and the switching between the wells. The variation of
$P_{1}$ can be used to characterize the stochastic resonance
phenomenon. For values of $D$ nearly above $D_{\mathrm{c}}$,
$\tau_{R}/T_{0}$ is distributed relatively over wide interval of
time (fig.10a). As $D$ increases the range of $\tau_{R}$ decreases
and  $P$ of smaller $\tau_{R}$ increases. At $ D =
D_{\mathrm{max}}$ as shown in fig.10c the noise induced
oscillations have period $ \approx T_{0}/2$ and hence
$P(\tau_{R}/T_{0} = 1/2)$ becomes maximum. As $D$ is further
increased $P_{1}$ decreases. Similar behaviour is observed for the
periodic forces given in figs.1a-d. Figure 11 shows $P_{1}$ versus
$D$ for various forces. The effect of various forces on $P_{1}$
can be clearly seen in this figure. For small values of $D$,
$\tau_{R}/T_{0}$ is distributed over wide interval of time.
Therefore, $P_{1}$ increases with increase in $D$, reaches a
maximum at $ D = D_{\mathrm{max}}$ and then decreases. With the
modulus of sine and rectified sine waves, the peak $P_{1}$ at
$\tau_{R}/T_{0} = 1/2 $ is numerically calculated for various
values of $D$ in the wells $V_{-+}$ and $V_{++}$. For the well
$V_{-+}$, the value of $P_{1}$ decreases with increase in the
value of $D$. But for the well $V_{++}$, $P_{1}$ increases with
increase in $D$ reaches a maximum at which $SNR$ is maximum and
then decreases. Figures 12 and 13 show the plot of $P_{1}$ versus
noise intensity $D$ for the forces in fig.1e and fig.1f
respectively.
 \vskip 10pt
  \noindent{\bf{5. Summary and conclusion}}
 \vskip 10 pt
     We have studied numerically the effect of different periodic
 forces and Gaussian noise in the overdamped two coupled
 anharmonic oscillators eqs(1). In order to compare the effect
 of external forces we have fixed the period of the forces as $2 \pi/ \omega$. The
 average of all the forces over one drive cycle is zero except for
 modulus of sine wave and rectified sine wave. The system shows
 periodic behaviour in the presence of external periodic forces
 and in the absence of noise term. Since the forcing is added only
 to $x$-component, the system performs transitions between the
 wells $V_{++}$ and $V_{-+}$ only. The value of $f_{\mathrm{c}}$
 is found to be small for square wave when compared with other
 forces. Cross-well behaviour is delayed in the case of sine,
 symmetric saw-tooth and asymmetric saw-tooth forces. With the
 other two forces, cross-well motion is not observed.  Stochastic
 resonance is observed with all the forces used. The mean residence time, power
 spectrum, $SNR$ and distribution of normalized residence times are
 used to characterize stochastic resonance. The features of
 stochastic resonance are found to be same for the forces such as sine, square,
 symmetric saw-tooth and asymmetric saw-tooth waves. Maximum $SNR$ is obtained at
 the value of $D$ at which sum of the residence times $ \approx \pi/ \omega $ for modulus
 of sine and rectified sine waves respectively. The values of $D_{\mathrm{c}}$
 and $D_{\mathrm{max}}$ are found to be different for different periodic forces.
 Signal-to-noise ratio values are found to depend on the type of forcing used. Square
 wave gives maximum $SNR$. Modulus of sine wave and rectified sine wave yields minimum $SNR$.
 The height of the peak $P_{1}$ at $D_{\mathrm{max}}$  obtained
 from normalized residence time distribution vary with the
 periodic forces. In the nonlinear dynamics literature, many
 interesting results have been obtained with the quasiperiodic,
 chaotic, aperiodic forces and with the different noise terms.
 Therefore, it is interesting to study the stochastic resonance
 behaviour with the above forces.
 \vskip 10pt
\noindent{\bf{Acknowledgement}}
 \vskip 10pt
The work reported here forms part of a Department of Science and
Technology, Government of India research project. We are thankful
to Dr.K.P.N. Murthy for stimulating discussion.


\newpage
\begin{figure}
\begin{center}
\epsfig{figure=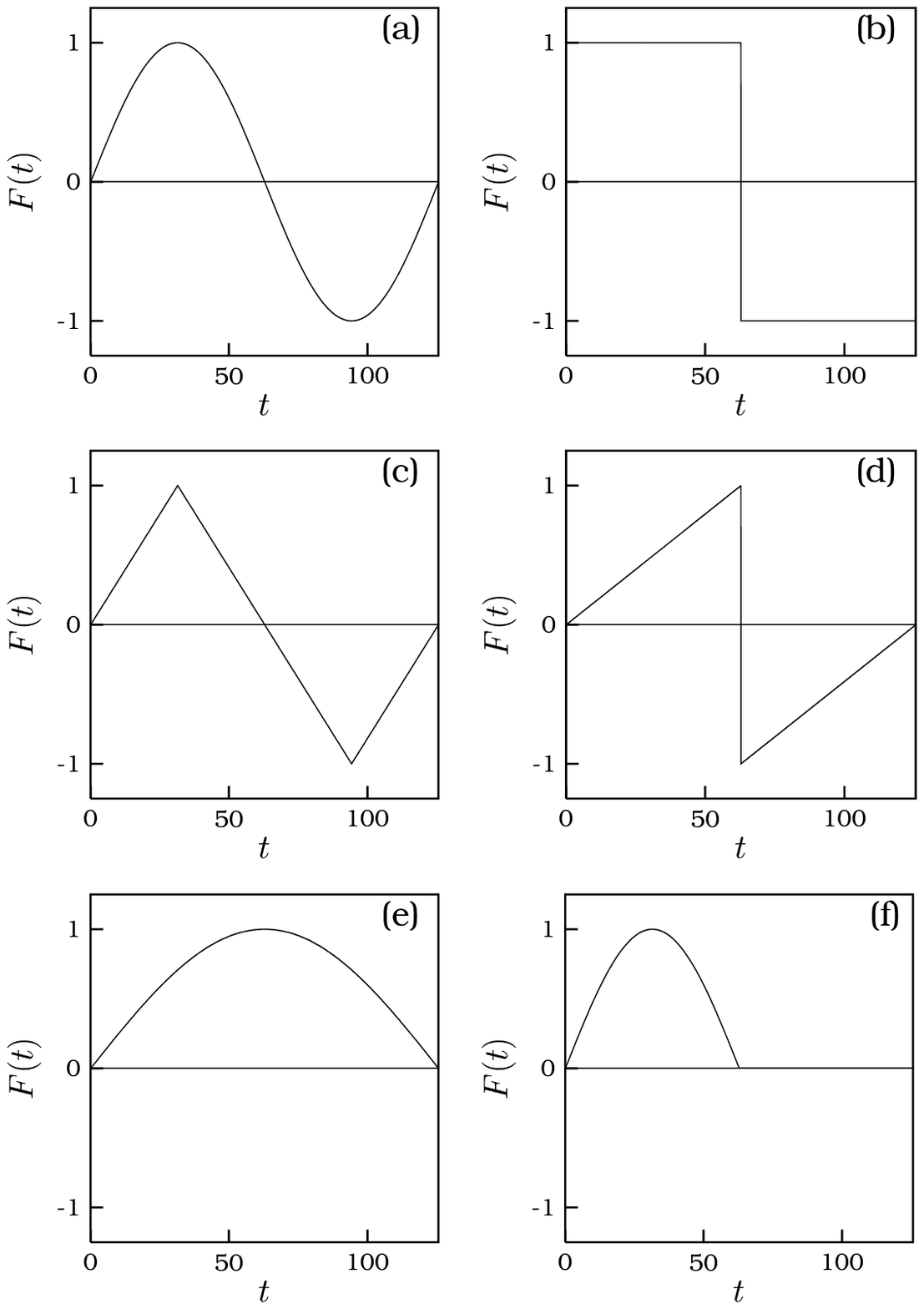, width=0.8\columnwidth}
\end{center}
\caption{Form of different periodic forces: (a) sine wave, (b)
square wave, (c) symmetric saw-tooth wave, (d) asymmetric
saw-tooth wave, (e) modulus of sine wave and (f) rectified sine
wave. For all the forces the period $T_{0}$ is $2 \pi/ \omega$
with $ \omega = 0.05$ and the amplitude is set to 1.} \label{Fig1}
\end{figure}
\begin{figure}
\begin{center}
\epsfig{figure=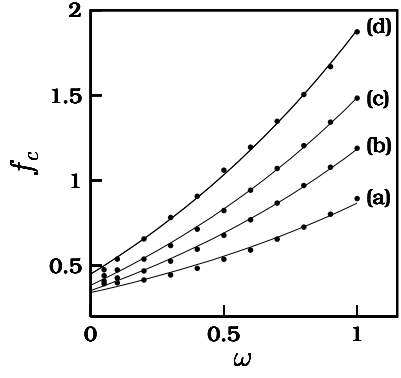, width=0.3\columnwidth}
\end{center}
\caption{$f_{\mathrm{c}}$ versus $ \omega$ for various forces. The
curves (a), (b), (c) and (d) are for the square wave, sine wave,
symmetric saw-tooth wave and asymmetric saw-tooth wave
respectively. Dots represent numerical data and continuous lines
are the best curve fits.}
 \label{Fig2}
\end{figure}
\begin{figure}
\begin{center}
\epsfig{figure=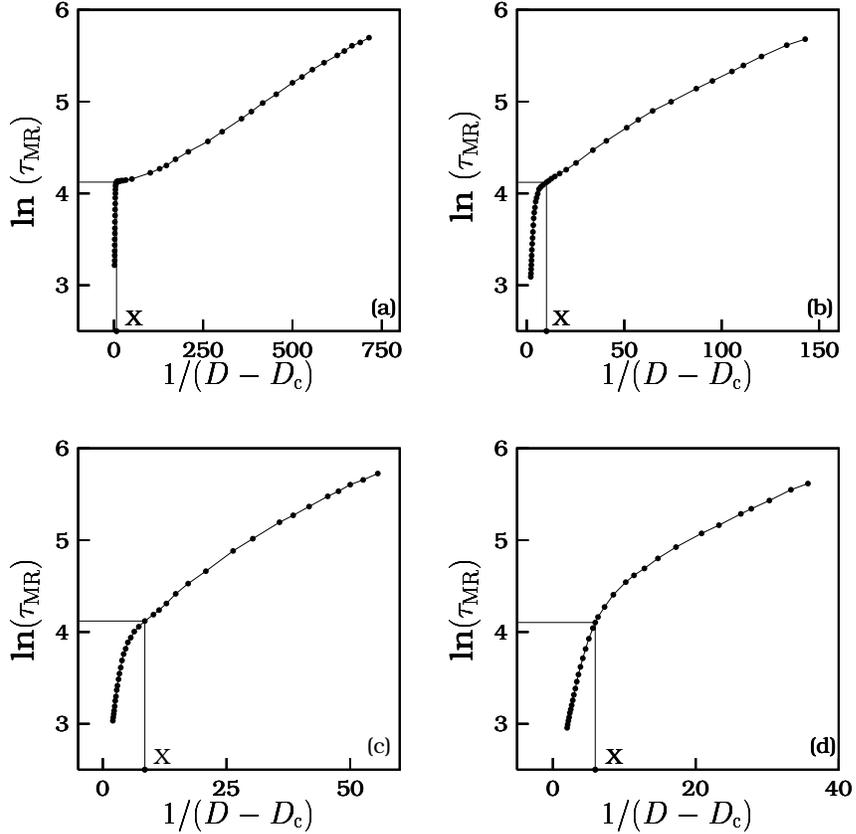, width=0.7\columnwidth}
\end{center}
\caption{Mean residence time in logarithmic scale versus
$1/(D-D_{c})$ for (a) square wave, (b) sine wave, (c) symmetric
saw-tooth wave and (d) asymmetric saw-tooth wave. The parameters
of the system are $ a_1 = 1.0$, $a_2 = 1.1$, $b_1 = 1.0$, $b_2 =
1.0$, $\delta = 0.01$, $f = 0.38$ and $\omega = 0.05$. The symbol
X indicates the value of $ 1/(D_{\mathrm{max}} - D_{\mathrm{c}})$
where $D_{\mathrm{max}}$ is the value of $D$ at which $ \tau_{MR}
= {\pi}/{\omega}$.}
 \label{Fig3}
\end{figure}

\begin{figure}
\begin{center}
\epsfig{figure=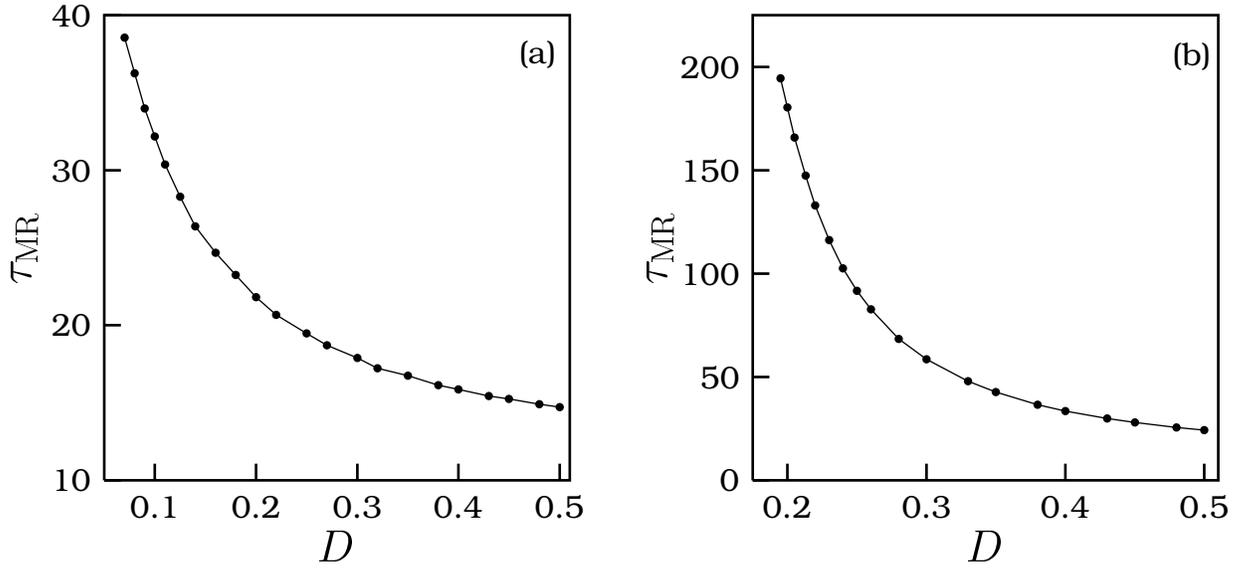, width=1.0\columnwidth}
\end{center}
\caption{Mean residence time in the wells (a) $V_{-+}$ and (b)
$V_{++}$ for the force modulus of sine wave. In the subplot (b)
for clarity $ \tau_{MR}$ is plotted from $D=0.2$.}
 \label{Fig4}
\end{figure}
\begin{figure}
\begin{center}
\epsfig{figure=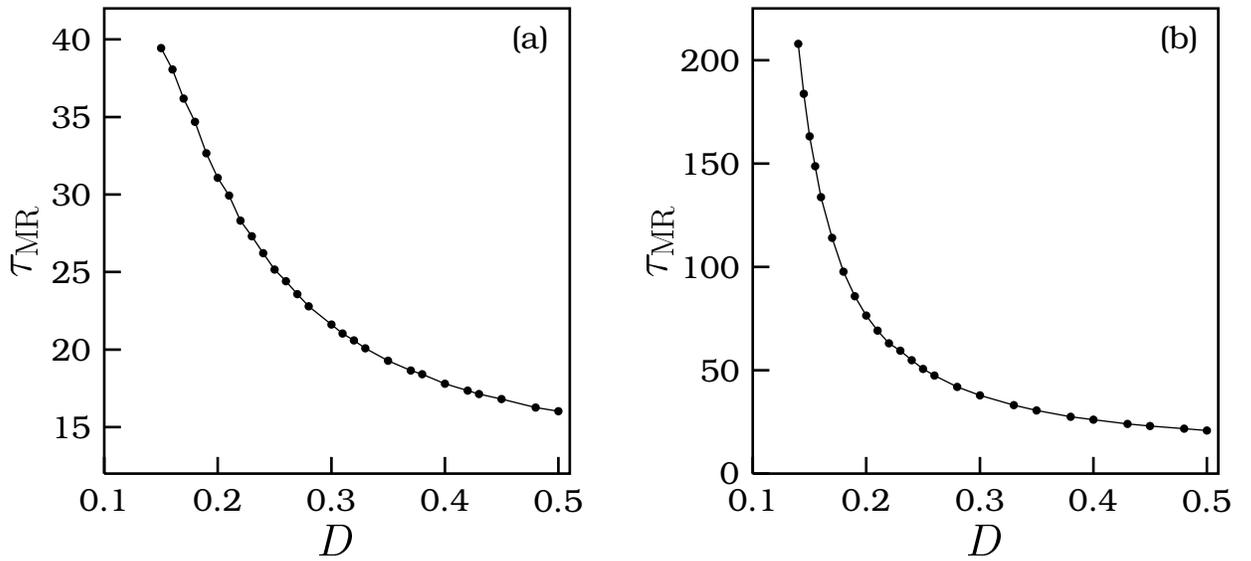, width=1.0\columnwidth}
\end{center}
\caption{Mean residence time of system (1) in the wells (a)
$V_{-+}$ and (b) $V_{++}$ for the rectified sine wave.}
 \label{Fig5}
\end{figure}
\newpage
\begin{figure}
\begin{center}
\epsfig{figure=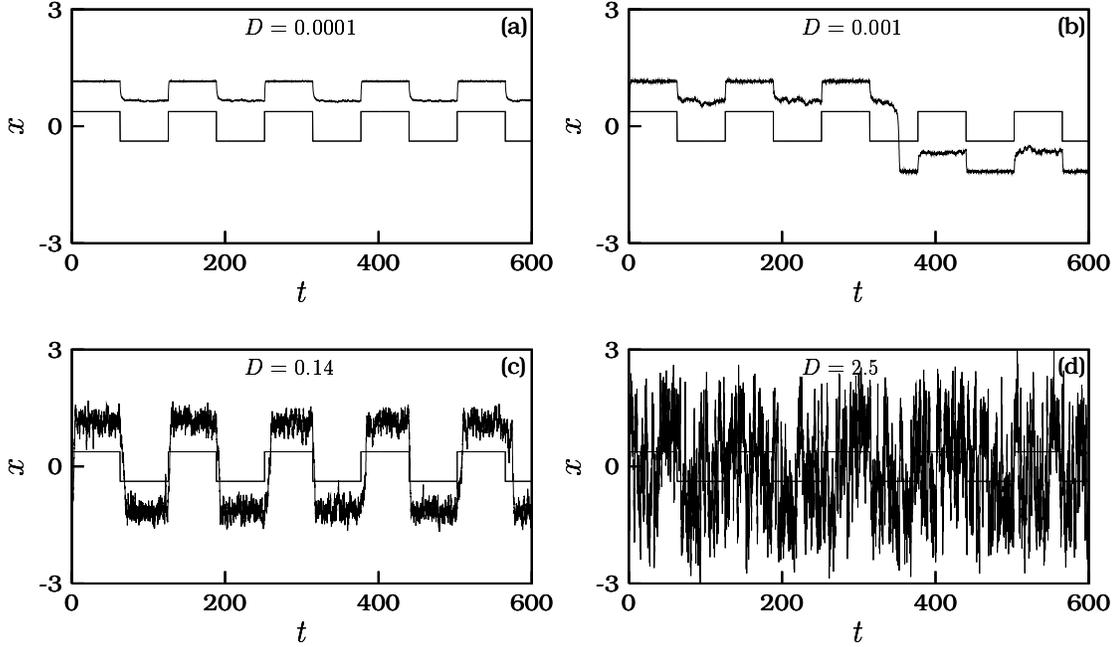, width=0.9\columnwidth}
\end{center}
\caption{Time series plot of eqs(1) in the presence of square wave
for few values of noise intensity $D$. The parameters of the
system are $ a_1 = 1.0$, $a_2 = 1.1$, $b_1 = 1.0$, $b_2 = 1.0$,
$\delta = 0.01$, $f = 0.38$ and $\omega = 0.05$. The applied
square wave is also plotted.}
 \label{Fig6}
\end{figure}
\begin{figure}
\begin{center}
\epsfig{figure=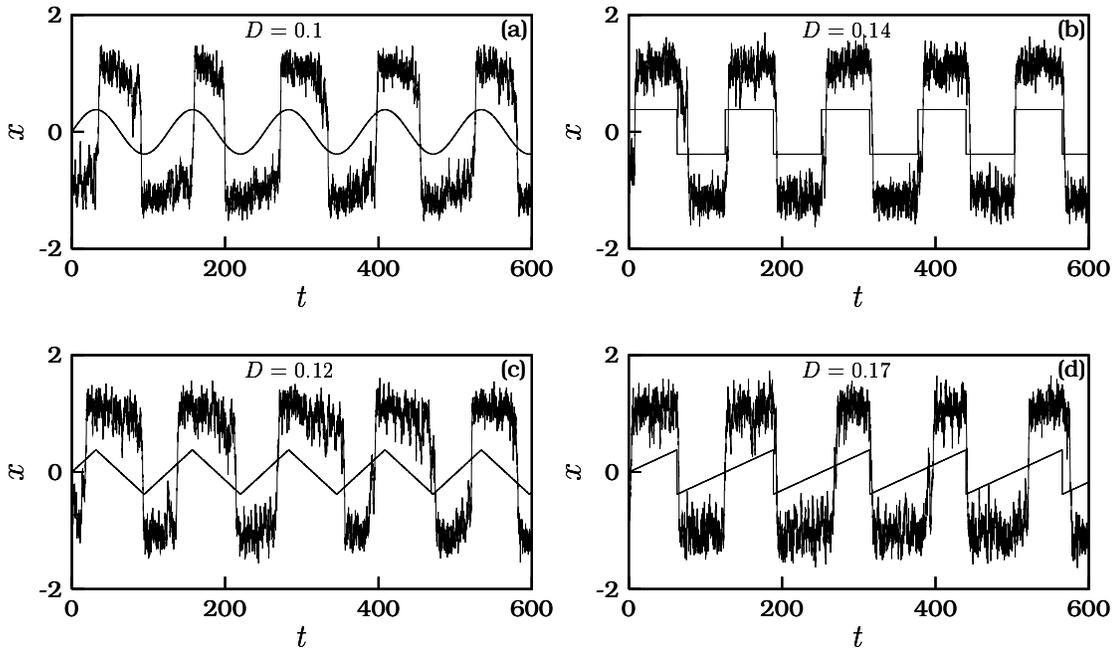, width=0.9\columnwidth}
\end{center}
\caption{Synchronization of the output signal with the different
applied periodic forces such as (a) sine wave, (b) square wave,
(c) symmetric saw-tooth wave and (d) asymmetric saw-tooth wave. In
all the subplots the value of $D$ is $D_{\mathrm{max}}$.}
 \label{Fig7}
\end{figure}
\begin{figure}
\begin{center}
\epsfig{figure=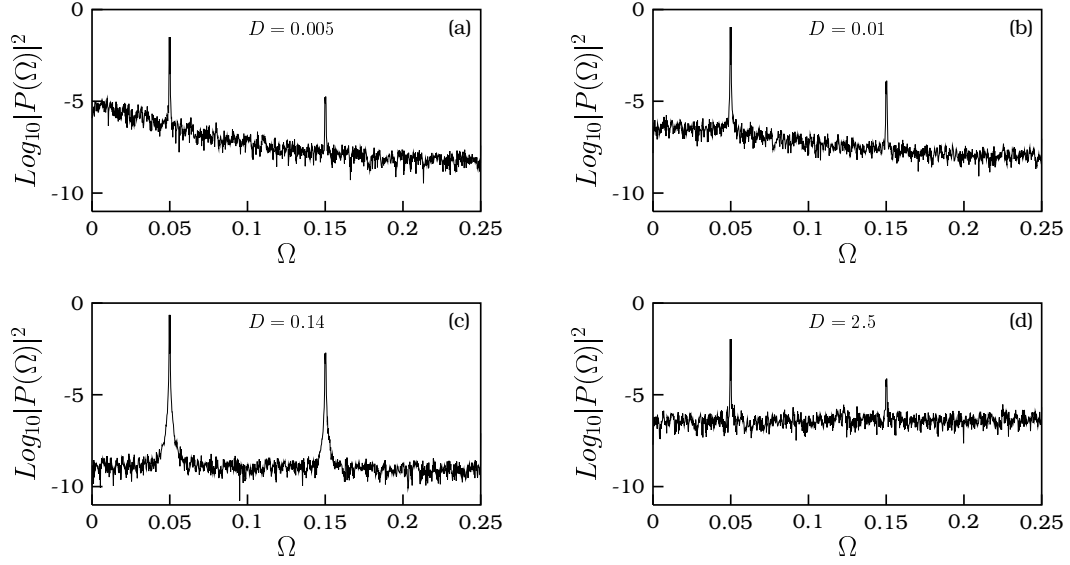, width=0.9\columnwidth}
\end{center}
\caption{Power spectral density plot in the presence of square
wave for few values of noise intensity $D$.}
 \label{Fig8}
\end{figure}
\begin{figure}
\begin{center}
\epsfig{figure=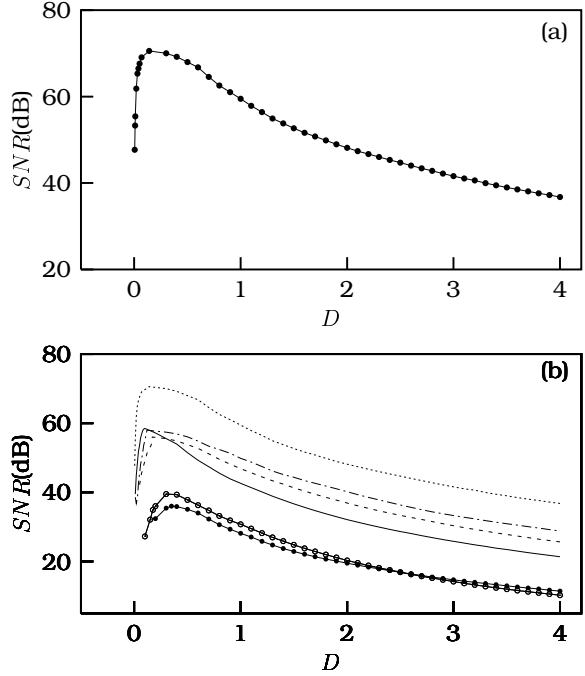, width=0.5\columnwidth}
\end{center}
\caption{ (a) Signal-to-noise ratio plot for range of values of
noise intensity $D$ in the presence of square wave. (b)
Signal-to-noise ratio plot for various periodic forces such as
sine (continuous curve), square (dotted curve), symmetric
saw-tooth ($- ${\huge{$_\cdot$}}$ -$ curve), asymmetric saw-tooth
(dashed curve),  modulus of sine (dots joined by a continuous
curve) and rectified sine (circles joined by a continuous curve)
used. In all the cases $f = 0.38$ and $ \omega = 0.05$.}
 \label{Fig9}
\end{figure}
\begin{figure}
\begin{center}
\epsfig{figure=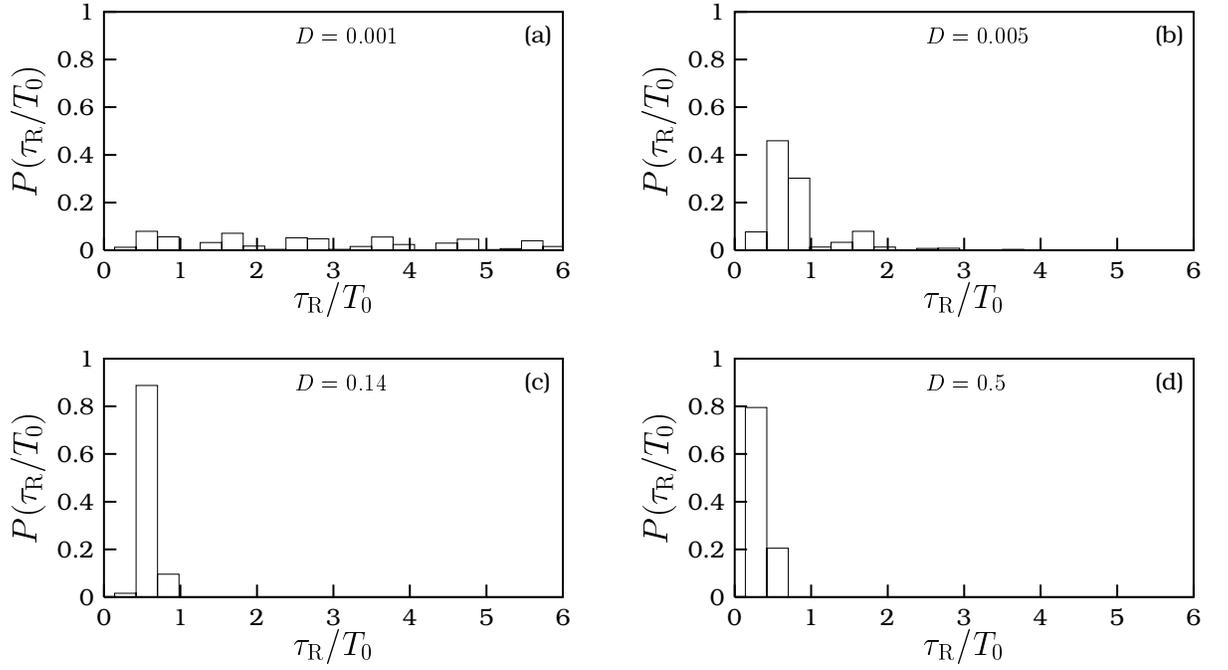, width=1.0\columnwidth}
\end{center}
\caption{Normalized residence time distribution for few values of
$D$ in the presence of square wave with amplitude $f=0.38$ and
frequency $ \omega = 0.05$.}
 \label{Fig10}
\end{figure}
\begin{figure}
\begin{center}
\epsfig{figure=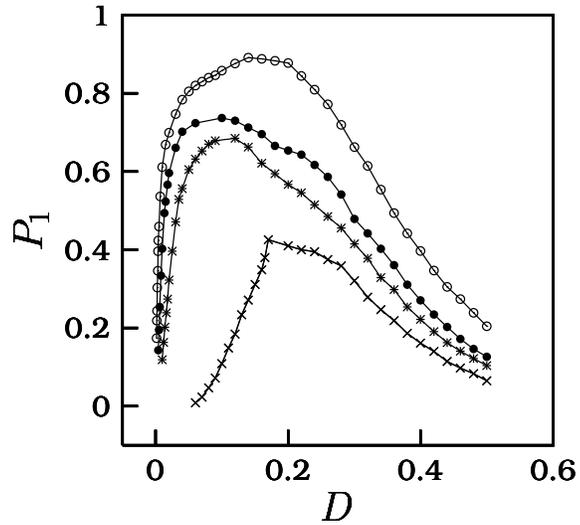, width=0.45\columnwidth}
\end{center}
\caption{Height of the peak $P_{1}$ in the normalized residence
time distribution at $ \tau_{R} = {T_{0}}/2$ for square wave
(marked by $\bullet$), sine wave $(\circ)$, symmetric saw-tooth
wave $(\ast)$ and asymmetric saw-tooth wave $(\times)$.}
 \label{Fig11}
\end{figure}
\begin{figure}
\begin{center}
\epsfig{figure=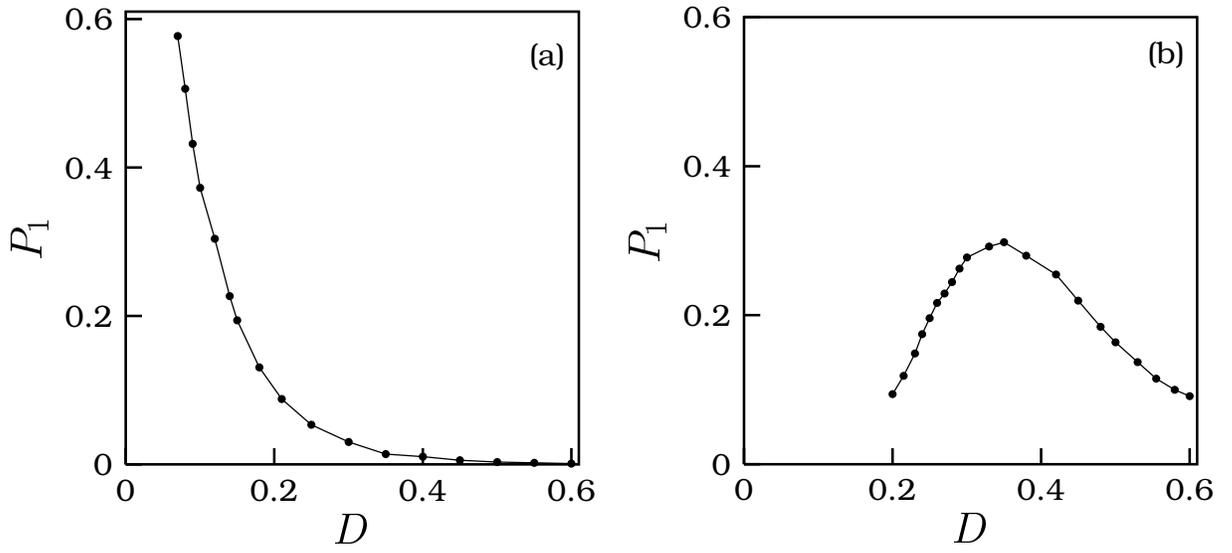, width=1.0\columnwidth}
\end{center}
\caption{ $P_{1}$ versus $D$ for the wells (a) $V_{-+}$ and (b)
$V_{++}$ with the applied force as modulus of sine wave.}
 \label{Fig12}
\end{figure}
\begin{figure}
\begin{center}
\epsfig{figure=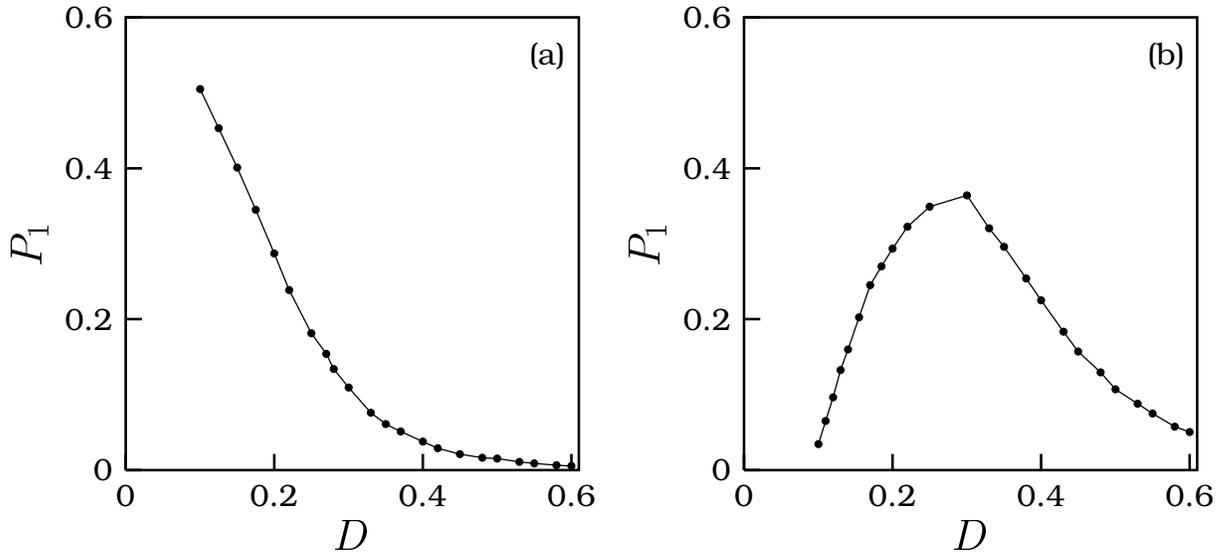, width=1.0\columnwidth}
\end{center}
\caption{$P_{1}$ versus $D$ for the wells (a) $V_{-+}$ and (b)
$V_{++}$ with the applied force as rectified sine wave.}
 \label{Fig13}
\end{figure}
\end{document}